# All-dielectric metasurface cylindrical lens

Jeongho Ha,[1,†] Abdoulaye Ndao,[1,†] Liyi Hsu[1], Jun-hee Park[1], and Boubacar Kanté[1,*]

[1]Department of Electrical and Computer Engineering, University of California San Diego, La Jolla, California 92093-0407, USA
[†]These authors contributed equally to this work
*Corresponding author: bkante@ucsd.edu

**Abstract**

*Conventional optical components have been proposed to realize high-quality line focusing with uniform intensity distribution such as cylindrical lenses, segmented wedge-arrays, or a combination of prisms and spherical mirrors. However, numerous factors such as the manufacturing tolerances of conventional lenses or the need for precise alignment of the lenses cause wavefront aberrations that impact the performance of optical systems. These aforementioned limitations of conventional optical components affect the uniformity of the intensity distribution. Here, we numerically and experimentally demonstrate an integrable planar all-dielectric cylindrical lens for uniform line focusing. The lens has a NA of 0.247 and a measured uniformity of 0.92% at 800 nm.*

**Introduction**

In recent years, ongoing efforts to miniaturize and increase the performance of microscale sized optical lenses have allowed the development of different imaging systems for the characterization of complex media. These systems, now accessible on almost all commercial spectrometers, have opened new fields of investigation, making it easy to characterize inhomogeneous samples. One can probe into two or even three-dimensional objects of macroscopic size (at the scale of mm or cm), while keeping a spatial resolution of the order of a micrometer, limited mainly by diffraction. Cylindrical lenses [1], random phase plates [2], segmented wedge arrays [3], and scanning mirrors [4] are often proposed as line focus generators for applications such as Raman spectroscopy [5] or X-rays imaging [6]. However, it is challenging to obtain simultaneously line focusing and high uniformity intensity distribution from integrated lenses.

Metasurfaces have been investigated as potential alternatives for integrated optical free space components [9-13]. Metasurfaces are subwavelength nanostructured devices that enable the control of optical wavefronts, polarization, and phase to build a large variety of flat optical components, including planar lenses [14-18], quarter-wave plates [19], half wave plates [20] optical vortex plates [21], carpet cloaks [22-23], solar concentrators [24-26], polarizers [27], thin absorbers [28], biomedical imaging devices [29] or sensors [30]. However, all these optical components are so far based on point-focusing. Here, we experimentally demonstrate a CMOS compatible and integrable planar all dielectric cylindrical lens for line-focusing. This lens generates line focusing with uniform intensity. The line-focusing planar lens paves the way to applications such hyperspectral Raman spectroscopy, line-focusing solar collectors, or wave probe with line focusing.

## Numerical simulation

To achieve the desired line-focusing at λ=800 nm, the phase profile of the wave front as a function of position $x$ along the metasurface lens must satisfy the parabolic equation [25]:

$$\Phi = k_0(\sqrt{x^2 + f^2} - f) \qquad (1)$$

where $k_0$ is the free space wave-vector, $x$ is the distance from the considered element to the center of the metasurface lens and $f$ is the focal length. The metasurface is made of $TiO_2$ cylinders (brown color) at 800 nm wavelength.

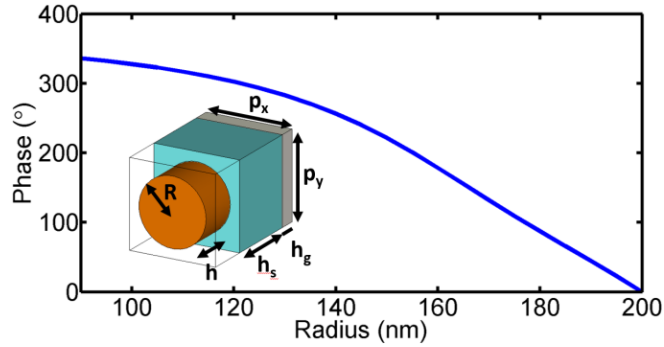

*Figure 1 Phase shift for different radii and schematic of the unit cylindrical elements. Dimensions are indicated on the figure and p is the period, h is the thickness of the resonator, R is the radius, $h_s$ is thickness of the spacer and $h_g$ the thickness of the ground plane. Materials: $TiO_2$ (brown), $SiO_2$ (blue), silver ground plane (grey).*

The inset of figure 1 shows the dielectric metasurfaces unit cell. It consists of a $TiO_2$ cylindrical resonator on a $SiO_2$ ($h_s$ = 440 nm)–silver ($h_g$ = 100 nm) multilayer. Silicon dioxide ($SiO_2$) is the spacer and silver (Ag) [31] serves as ground plane for the lens to work in reflection. The thickness ($h$) of the cylindrical resonators is fixed to 250 nm and the radius ($R$) changes from 90 nm to 200 nm. The metasurface is periodic along the y direction with a sub-wavelength unit cell ($p_x = p_y$ = 510 nm). Due to the low loss of glass and $TiO_2$ at the desired frequency (800 nm), the metasurface is considered lossless. Using ellipsometry, the refractive indices of $TiO_2$ and the $SiO_2$ are measured and found to be ≈2.325 and ≈1.49 respectively (average of 5 measurements). To obtain the required 2π phase-shift (see Fig. 1), the scattering coefficients of the resonators are calculated using the commercial software CST, and, boundary conditions in x and y directions are set to periodic. Figure 1 shows that the 2π phase-shift is achieved by changing the radius from 90 nm to 200 nm. To prove the feasibility of the device, we designed a cylindrical metasurface lens with a focal length 40 µm and a size of 20 µm by 20 µm. Using equation 1, the required phases are calculated. By interpolating Fig. 1, the radius of each element fitting the required phase is obtained (see Fig. 2).

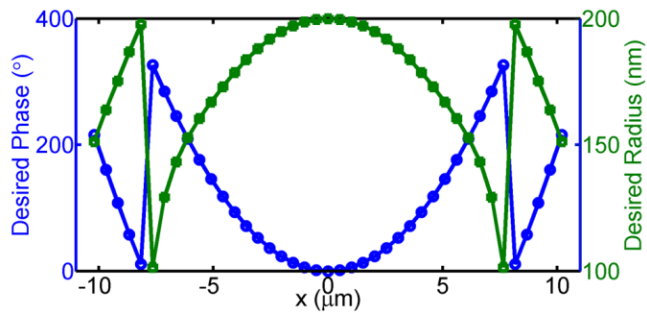

*Figure 2 Phase distribution of a planar cylindrical lens under normal incidence with a 40 µm focal length and a 20 µm by 20 µm aperture size (blue). Required radius (green).*

Figure 2 presents the required phases and the corresponding radii of resonators. To have a line focus along the *y* direction, elements in the y-direction are identical with a periodicity $p_y$.

**Fabrication and measurement**

Figure 3 illustrates the fabrication flowchart of the structures using top-down etching processes. The all-dielectric planar cylindrical lens is fabricated on a fused silica substrate. It should be noted that the substrate always has a thin layer of native oxide, which has a crystallographic mesh parameter incompatible with that of gold or silver, resulting in a quasi-zero adhesion to the substrate. Therefore, we deposit, using electron beam metal deposition, a thin layer (5 nm) of Germanium (Ge) on the substrate to overcome the lack of adhesion since Ge improves adhesion and surface smoothness. A silver metal layer of 100 nm (larger than the skin depth of light in the visible) is then deposited without opening the chamber by changing the crucible [Fig. 3(a)]. Then, a 440 nm thick $SiO_2$ layer was deposited on the Ag layer using PECVD (Plasma-enhanced chemical vapor deposition) as shown in [Fig. 3(b)]. An electron beam resist (PMMA), is spun on the sample followed by baking at 170 °C (oven)). The metasurfaces pattern is written in the resist using electron beam lithography (EBL) [Fig. 3(c)] and subsequently developed in solution to remove the exposed e-beam resist (EBR). This pattern is the inverse of our final metasurfaces. The exposed sample is transferred to an atomic layer deposition (ALD) chamber producing amorphous $TiO_2$ while avoiding the contamination of the ALD chamber by the EBR. The ALD process ($TiCl_4$ precursor) deposits 250 nm of $TiO_2$ so that all features are filled with $TiO_2$ [Fig. 3(d)]. The residual $TiO_2$ film that coats the top surface of the resist is removed by RIE (reactive-ion-etching) process [Fig. 3(e)]. After removing the PMMA based structure, periodic $TiO_2$ metasurfaces were obtained [Fig. 3(f)].

To characterize the cylindrical lenses, sample sizes of 200 μm by 200 μm were fabricated.

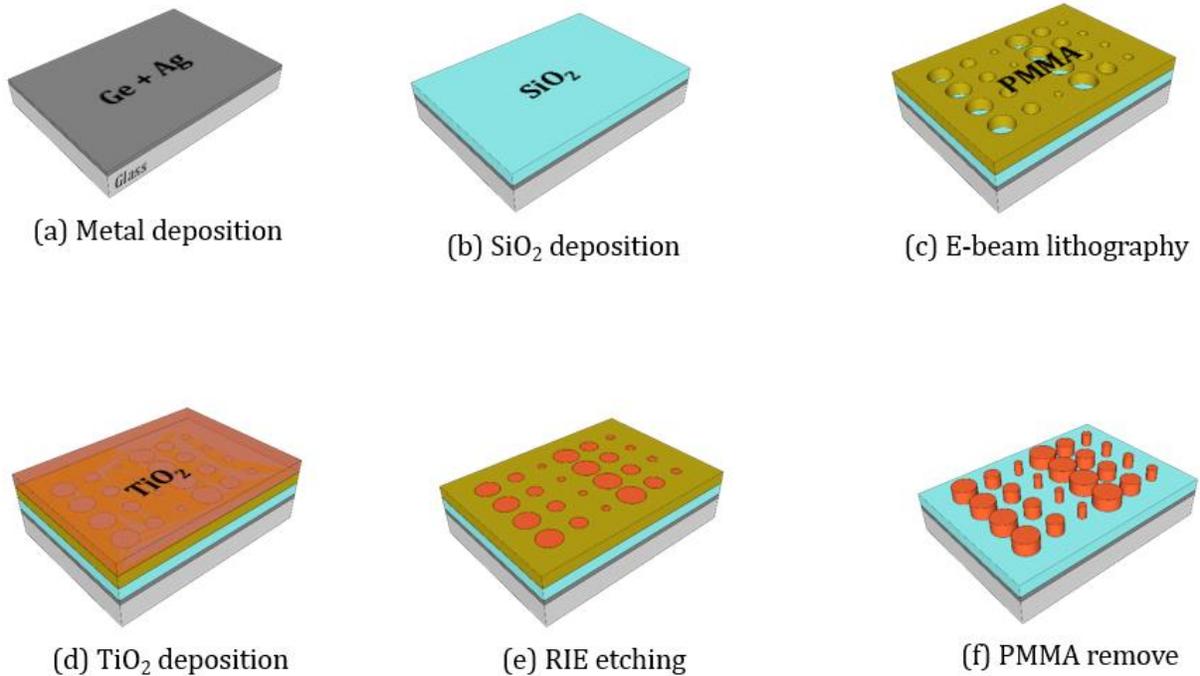

*Figure 3 Fabrication process: (a) metal deposition using E-beam evaporation system on the cleaned BK7 glass substrate, (b) $SiO_2$ deposition using PECVD, (c) E-beam lithography for gradient cylinder pattern using PMMA with 250 nm thickness (d) Thin layer $TiO_2$ deposition using ALD, (e) $TiO_2$ RIE process to expose the underlying PMMA pattern, (f) removal of the PMMA residues to obtain metasurface structures.*

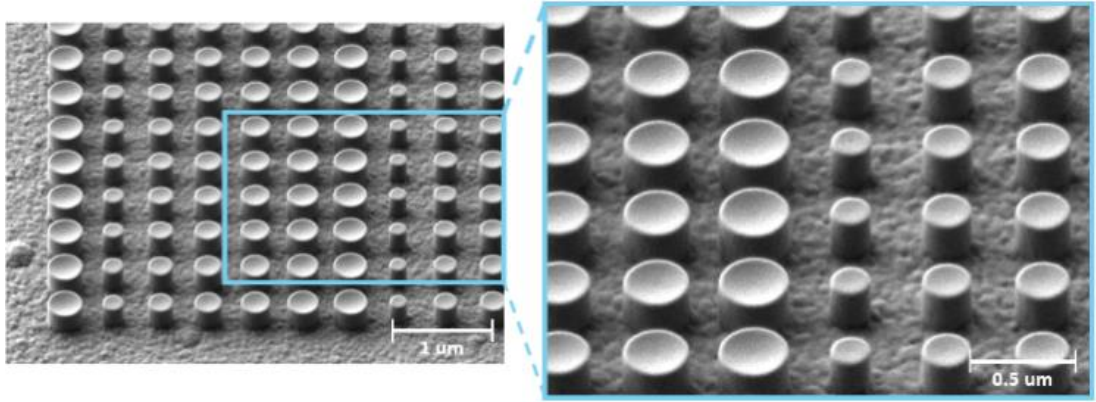

*Figure 4 (a) Top view Scanning electron microscope images of fabricated TiO$_2$ metasurfaces, (b) zoom-in containing 6 x 5 cylindrical structures imaged in the xy plane and clearly evidencing the gradient in structures size.*

Scanning electron microscope (SEM) images, presented in Fig. 4, shows successful fabrication of the structures with good quality. The geometrical parameters of the samples are compared with the theoretical specifications (nominal dimensions) in table 1. The measured geometrical parameters are slightly different from the nominal ones but are within ±5 nm confirming the good quality of the samples.

**Table 1. Comparison of nominal geometrical parameters and measured geometrical parameters.**

|  | Radius (nm) | | | | | |
|---|---|---|---|---|---|---|
| **Nominal values** | 200 | 120 | 150 | 165 | 170 | 180 |
| **Measured values** | 200 | 119 | 143 | 157 | 169 | 182 |

To optically characterize the focusing capability of the fabricated lenses, a custom setup presented in Fig. 5 was used. The experimental setup is composed of two main systems dedicated to illumination and imaging. The illumination system comprises a supercontinuum laser (NKT photonics) and an acousto-optic tunable filter (Super K) to select the operating wavelength (800 nm). A beam at $\lambda$ = 800 nm was transmitted through a broadband polarizing beam (50:50) splitter to illuminate the lens. The focused reflected beam is collected by the imaging system via the beam splitter. The imaging system consists of an extra-long working distance objective (x50), a corresponding lens, and a camera. Due to the limited space between the objective and the lens, an extra-long working distance (WD = 10 mm) was used. The lens was also mounted on a stage to adjust the distance between the lens and objective (50× objective).

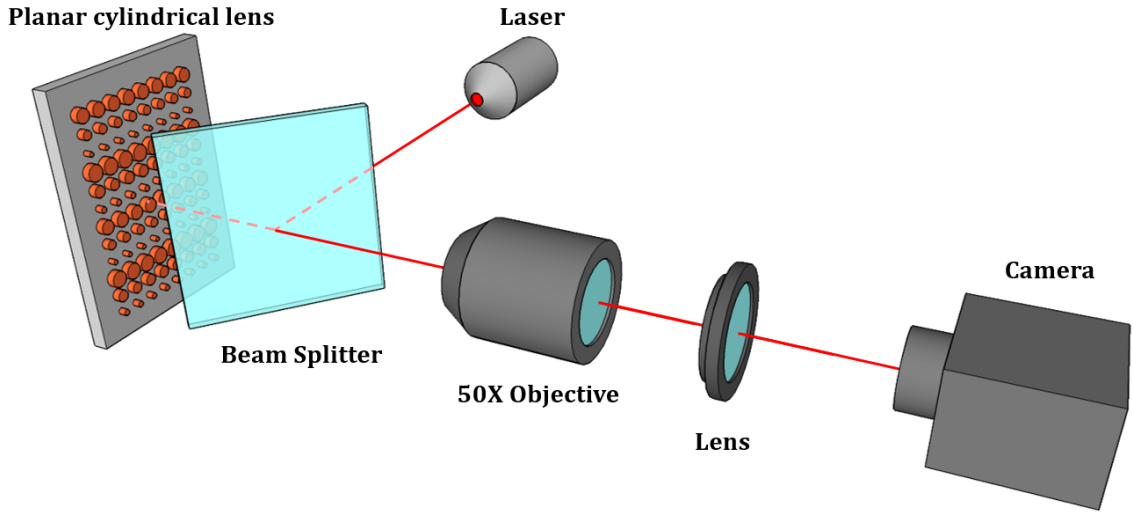

*Figure 5 Experimental setup for optical characterization of the planar cylindrical lens.*

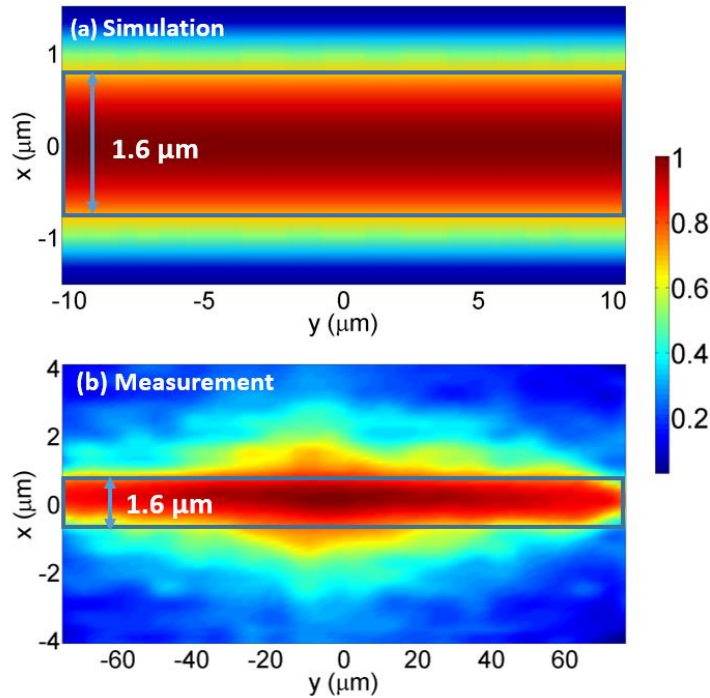

*Figure 6 Intensity profile of the reflected beam by the lens at 800 nm: (a) Numerical simulation of line-focusing using the structured metasurface, (b) Experimental measurement of line-focusing for the fabricated planar cylindrical lens.*

Figure 6 presents the numerical simulation [Fig. 6(a)] and experimental [Fig. 6(b)] normalized electrical field intensity at the focal spot in the *xy* plane. A clear and uniform line focus is observed. The measured intensity profile illustrates the line focus at the focal distance (0.4 mm). To better quantify the uniformity of the focus, we first calculated the standard deviation of the electric field intensity ($\sigma$) at the focal line (along y). The uniformity is then 1 - $\sigma$ meaning that the uniformity is unity when there is no deviation (perfect uniformity). However, the focus is not a perfect line but rather an area with a width limited by diffraction. In our design at the wavelength of 800 nm, the numerical aperture (NA) is 0.247, and the diffraction limit is thus $d = 1.6$ μm. The average of the field uniformity along the *x*-direction in a 1.6 μm wide region centered at the focal line gives a quantitative comparison of the focusing in experiment and simulation. The experimental result (0.92%) [Fig. 6(b)] agrees well with numerical simulation (0.99%) [Fig. 6(a)] although the measured uniformity in experiment is 6 % smaller than the numerical simulation uniformity (0.99%). The small discrepancies between numerical simulations and experimental results are mainly attributed to the fabrication imperfections inducing slightly non-uniformity in experimental results and the non-fully planar wavefronts incident on the samples in experiments. Indeed, the incident wave on the sample has a Gaussian profile with higher intensity at its center resulting in a stronger field at the center of the lens as observed experimentally in Fig. 6 (b).

**Conclusion**

In summary, we experimentally demonstrated a planar all-dielectric cylindrical lens for line focusing. The lens has a NA of 0.247 and measured uniformity 0.92% at 800 nm. This lens not only provides line focusing but also uniform intensity distribution along a line with a width mainly limited by diffraction. This planar cylindrical lens paves the way to applications such as high-resolution Raman spectroscopy, line-focusing solar collectors, or wave probe with line focusing.